\documentclass[sn-basic]{sn-jnl}% Basic Springer Nature Reference Style/Chemistry Reference Style
\usepackage{graphicx}%
\usepackage{multirow}%
\usepackage{amsmath,amssymb,amsfonts}%
\usepackage{amsthm}%
\usepackage{mathrsfs}%
\usepackage[title]{appendix}%
\usepackage{xcolor}%
\usepackage{textcomp}%
\usepackage{manyfoot}%
\usepackage{booktabs}%
\usepackage{algorithm}%
\usepackage{algorithmicx}%
\usepackage{algpseudocode}%
\usepackage{listings}%
\usepackage[T1]{fontenc}

\begin{document}

\title[Article Title]{The early Solar System and its meteoritical witnesses}


\author*[1]{\fnm{Emmanuel} \sur{Jacquet}}\email{emmanuel.jacquet@mnhn.fr}


\author*[2]{\fnm{Cornelis} \sur{Dullemond}}\email{dullemond@uni-heidelberg.de}

\author[3]{\fnm{Joanna} \sur{Dr\k{a}\.{z}kowska}}\email{drazkowska@mps.mpg.de}

\author[4]{\fnm{Steven} \sur{Desch}}\email{steven.desch@asu.edu}

\affil*[1]{\orgdiv{IMPMC}, \orgname{Mus\'{e}um national d'Histoire naturelle}, \orgaddress{\street{57 rue Cuvier}, \postcode{75005}, \city{Paris}, \country{France}}}

\affil[2]{\orgdiv{Institute for Theoretical Astrophysics, Center for Astronomy}, \orgname{Heidelberg University}, \orgaddress{\street{Albert Ueberle Str. 2}, \city{Heidelberg}, \postcode{69120}, \country{Germany}}}

\affil[3]{\orgname{Max Planck Institute for Solar System Research}, \orgaddress{\street{Justus-von-Liebig-Weg 3}, \city{Göttingen}, \postcode{D-37077}, \country{Germany}}}

\affil[4]{\orgdiv{School of Earth and Space Exploration}, \orgname{Arizona State University}, \orgaddress{\street{PO Box 876004}, \city{Tempe}, \postcode{85287-6004}, \state{Arizona}, \country{USA}}}


\abstract{Meteorites, and in particular primitive meteorites (chondrites), are irreplaceable probes of the solar protoplanetary disk. We review their essential properties and endeavour to place them in astrophysical context. The earliest solar system solids, refractory inclusions, may have formed over the innermost au of the disk and have been transported outward by its expansion or turbulent diffusion. The age spread of chondrite components may be reconciled with the tendency of drag-induced radial drift if they were captured in pressure maxima, which may account for the non-carbonaceous/carbonaceous meteorite isotopic dichotomy. The solid/gas ratio around unity witnessed by chondrules, if interpreted as nebular (non-impact) products, suggests efficient radial concentration and settling at such locations, conducive to planetesimal formation by the streaming instability. The cause of the pressure bumps, e.g. Jupiter or condensation lines, remains to be ascertained.}




\maketitle

\include{abbrev}

\section{Introduction}\label{sec1}

  The early history of the Solar System is still largely a (space-time) \textit{Terra Incognita}. To be sure, the relevant physics is essentially classical, so were we to know the initial conditions of the protosun, its protoplanetary disk and its environment, and had we infinite computational power, the evolution would be predetermined. Yet we are far from being such Laplace daemons. Even assuming present-day protoplanetary disks are a suitable proxy of the solar nebula, telescopic observations could not resolve any substructure in these disks until about a decade ago. The Very Large Telescope (VLT) at Paranal in Chile now takes detailed images of protoplanetary disks around young stars about 140 parsec away, but cannot probe the disk midplanes owing to the optical thickness of the disks at optical and infrared wavelengths%, and is thus mainly sensitive to stellar photons scattered off micron-size dust grains lofted to the surface
. Still, the Atacama Large Millimeter Array at Llano de Chajnantor in Chile (ALMA) can observe these disks at millimeter wavelengths, which are dominated by direct thermal emission from large (approximately millimeter size) dust grains mostly settled toward the midplane, but the resolution remains limited to 10s of au at best. On the theoretical side, while computational power has greatly increased over recent decades, simulations can be carried out over a disk lifetime only if resolution is kept coarser than the spatial scales relevant to turbulence etc., so require assumptions on sub-grid physics. Some may be tested by more local simulations over shorter timescales (typically hundreds of orbits), others calculated \textit{ab initio} but the knowledge of effective physical laws and appropriate approximation regimes remains fragmentary.

  In this context, meteorites provide key constraints. Multiple lines of evidence indicate that they record the first few million years of existence of the Solar System, when the proto-Sun was still surrounded by its gaseous protoplanetary disk, also known as the "solar nebula". Save for rare Martian and lunar samples, their radiometric absolute ages tightly cluster around 4.56 Ga, consistent with the helioseismological dating at 4600$\pm$40 Ma \citep{HoudekGough2011}, or 4569$\pm$6 Ma or 4587$\pm$7 Ma \citep{BonannoFroehlich2015}.  The oldest Solar System-made meteoritic components, the refractory inclusions, set a Time Zero at 4567 or 4568 Ma \citep{BouvierWadhwa2010, Connellyetal2017,Pirallaetal2023,Deschetal2023}, notwithstanding a systematic error due to the uncertainty on the half-lives of uranium isotopes \citep[e.g.][]{Tissotetal2017}. The evidence of short-lived radionuclides such as aluminum-26 suggest a recent (more or less averaged) input from massive stars in the parental molecular cloud \citep[e.g.][]{Deschetal2022}. The composition of chondrite components, in particular refractory inclusions, bear witness to exchanges with a gas of solar (H$_2$-dominated) composition \citep{DavisRichter2014}, that is a protoplanetary disk environment, before disk accretion and photoevaporation dissipated it. Paleomagnetic data suggest the disappearance of nebular magnetic fields, and of the solar nebula altogether, by $\sim$4 Ma \citep{Weissetal2021}, consistent with the few-Ma disk half-lives in neighbouring molecular clouds \citep{WilliamsCieza2011}. Meteorites are thus invaluable time probes of disk interiors at optical depths and microscopic scales that will always elude telescopic capabilities.

  Still, meteorites are largely rocks without geological context. Even in those cases where a pre-encounter orbit can be calculated, the chaotic orbital evolutions undergone since ejection from their parent bodies preclude the ready identification of the latter. Even when the parent bodies are identified, where they orbit now is not necessarily a good proxy of where they originally accreted, depending on the orbital instabilities that reshuffled planet trajectories in the past \citep[e.g.][]{Walshetal2011}. As to chronology, radiometric age dating at sub-Ma resolution necessary to distinguish disk evolutionary periods remains challenging and somewhat model-dependent, if at all possible for the events of interest. Planetesimal accretion, for example, cannot be directly dated and is a free parameter of thermal evolution models. Also, what part of meteorite petrology relates to the "background" protoplanetary disk in contradistinction to more local (e.g. planetary) environments is not necessarily clear beforehand. Meteorites are thus not ideal sensors of nebular conditions at known positions and epochs. This is somewhat like trying to unravel the history of a destroyed Egyptian pyramid based on debris randomly scattered around the globe. Yet this is the beauty of it. The very astrophysics which meteorites are called to help is thus also needed to provide template contexts to make sense of them. Astrophysics constrain meteoritics as much as meteoritics constrain astrophysics \citep{Jacquet2014review}.

  The purpose of this opening chapter is thus to provide some basic astrophysical context on the early Solar System for the more specific meteorite-based discussions of the ensuing chapters. We will focus on the evolution of the solids in the disk. Yet, however astronomically-minded, these introductory sections could not, of necessity, be blind to meteoritical constraints to guide their very content, and we will thus start with cosmochemical prerequisites.

\section{Meteoritical prolegomena}

\subsection{Overview of meteorite classification}
\subsubsection{Petrology}
A first-order division among meteorites separates \textit{primitive meteorites}, or \textit{chondrites}, from \textit{differentiated meteorites}. Chondrites represent about 86 \% of falls \citep{Grady2000}. As supported by their close match with the solar photosphere for nonvolatile element abundances (especially for CI chondrites, see \textbf{Lodders chapter}), chondrites represent material largely unmodified since the accretion of their parent bodies. They are conglomerates of various solids (chondrules, refractory inclusions etc.) which were once freely floating in the solar nebula and/or in a planetesimal-forming environment and whose presentation is deferred until the next subsection. %For now, chondrites will be treated, as it were, as nebular "black boxes".

  Not all planetesimals retained a chondritic structure after accretion. The decay of short-lived radionuclides, chiefly aluminum-26, caused heating and partial melting of metal-sulfides (or magnetite for the most oxidized protoliths; \cite{McCoyetal2018}) and silicates in many of them. Metal-sulfide melts, driven by negative buoyancy, segregated into cores, while silicate partial melts migrated toward the surface to form a crust enriched in incompatible elements (that is, not easily accomodated in the crystal lattice of major rock-forming minerals), with or without a magma ocean stage. The intervening mantle comprises residues left over after melt extraction and/or early crystallizations from the said primary melts. The whole process leading to these compositionally distinctive layers in the parent bodies is called \textit{differentiation} \citep{McCoyetal2006}. Among differentiated meteorites, \textit{irons} are iron-nickel-dominated fragments arising from the core and \textit{achondrites} are the stony fragments of the mantle or crust. The latter comprise \textit{evolved achondrites}, which are igneous or magmatic rocks (i.e. rocks crystallized out of magmas) conformable to terrestrial terminology (basalts, orthopyroxenites, etc.; \cite{Jacquet2022}) and rarer \textit{primitive achondrites}, which are partial melting residues (or restites) and make a transition with chondrites%, with some occasionally containing relict chondrules from their protoliths
. Some irons and achondrites may however arise from localized impact melting on chondritic targets.

  Although the focus of most of this book will be on chondrites, it is worth noting that the parent bodies of differentiated meteorites must have accreted earlier than those of present-day chondrites, so as to secure enough radioactive heat for partial melting (even though the differentiated meteorites themselves may have crystallized and reached isotopic closure later). Hf-W dating of irons indeed suggest accretion within $\sim$1 Ma after refractory inclusions \citep{Spitzeretal2021} (see \textbf{Sch\"{o}nb\"{a}chler chapter}). Differentiated parent bodies represent thus an early generation of planetesimals unsampled by chondrites that must be taken into account in models of disk-wide compositional evolution.

 While heating was weaker for them, chondrites were not wholly spared by secondary processes (within the parent bodies). Melting of co-accreted ice led to aqueous alteration of some originally anhydrous minerals, especially in carbonaceous chondrites. Further temperature increases (e.g. in the drier ordinary chondrites) caused metamorphism, with various degrees of recrystallization and gradual disappearance of the original chondritic texture. \cite{VanSchmusWood1967} devised a scale of petrographic (or petrologic) types from 1 to 6 to quantify these secondary effects.   As reinterpreted in the 1970s, type 3 corresponds to the most primitive chondrites, while types 2 and 1 mark increasing degrees of aqueous alteration and types 4 to 6 increasing metamorphism%, which is not incompatible with some traces of aqueous and/or metasomatic alteration \citep[e.g.][]{Alexander1989,McCantaetal2008,Lewisetal2022}
. The later introduced types 7 \citep{Tomkinsetal2020} and 8  \citep{Jacquet2022} already correspond to primitive achondrites. Type 3 chondrites are not free of secondary effects and have been subdivided in decimal subtypes (e.g. 3.0, 3.1, 3.2 etc.; \cite{Searsetal1980,GrossmanBrearley2005,Sears2016}). Disentangling secondary from primary (preaccretionary) features is thus not always unequivocal (see \textbf{Krot, Lee chapters}).

\subsubsection{Genetics}
  Geological processes within asteroids are not the only cause of variation among meteorites. It had long been known that primitive meteorites differ in redox state \citep{Prior1920}, but \cite{UreyCraig1953} showed that they did in bulk composition too ("bulk", or "whole-rock" refers to the global composition of a rock, irrespective of the distribution of the elements among its petrographic components). The Urey-Craig work evinced metal-silicate (Fe/Si) fractionation (\textit{fractionation} meaning the partial or total separation of compositionally distinct fractions), but other fractionations were later identified, e.g. varying degrees of refractory element enrichment and/or depletions of volatile elements (see \textbf{Sossi chapter}) as well as Mg/Si fractionation \citep{Anders1964,LarimerAnders1970,WassonChou1974,Braukmuelleretal2018}. Isotopic ratios, starting with oxygen \citep{Claytonetal1973,Claytonetal1976}, were also later found to vary away from terrestrial values \citep{DauphasSchauble2016,Kleineetal2020}. Last not least, these whole-rock compositional variations correlate with variations in the textures and abundances of the chondrite components directly visible under the microscope, and it is an open question to what extent chondrite fractionations reflect varying proportions of a small number of component types common to all the disk \citep{Anders1964,Zandaetal2006,Alexander2019CC,Alexander2019NC,BrysonBrennecka2021} or not \citep{Goldbergetal2015,Hezeletal2018}. At any rate, obviously, the solar nebula was not homogeneous. It may have inherited heterogeneity from the parental molecular cloud but it also developed some \textit{in situ} (see \textbf{Burkhardt chapter}).

\begin{figure}[htbp]%
\centering
\includegraphics[width=\textwidth]{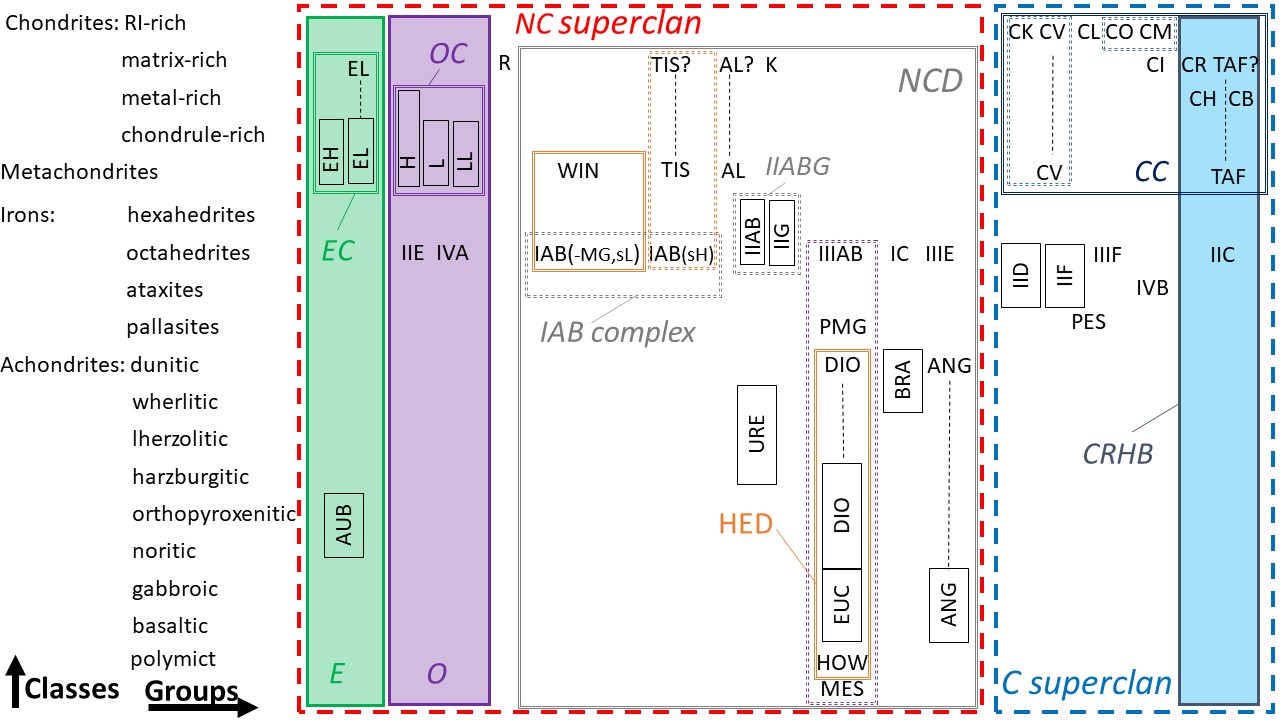}
\caption{Asteroidal meteorite taxonomy. The $y$ axis displays a petrologic classification of meteorites (that is, based on the nature of the rock; chondrites on top and differentiated meteorites at bottom) whereas the arbitrary $x$ axis allows the representation of genetic affinities (made explicit by boxing in high-level groups\textbf{; dashed lines also link different portions of low-level groups unrepresented in the intervening petrologic classes}). For example, aubrites (AUB), which are orthopyroxenitic achondrites, and EH and EL chondrites sample a common isotopic reservoir termed the E reservoir (colored in green), itself a subset of the NC superclan.  Abbreviations: AL= Acapulco-Lodran group (acapulcoites and lodranites), ANG = angrite, AUB = aubrite, BRA = brachinite, C = carbonaceous, CB = carbonaceous Bencubbin-type, CH = carbonaceous ALH 85085-type, CI = carbonaceous Ivuna-type, CK = carbonaceous Karoonda-type, CL = carbonaceous Loongana 001-type, CM = carbonaceous Mighei-type, CO = carbonaceous Ornans-type, CR = carbonaceous Renazzo-type, CV = carbonaceous Vigarano-type, DIO = diogenite, E = enstatite (EH, EL), H = high iron, HED = howardite-eucrite-diogenite, HOW = howardite, K = Kakangari-type, L = low iron, LL = low iron low metal, MES = mesosiderite, NC = noncarbonaceous, NCD = normal NC Differentiated meteorite signature, O = ordinary, PES = pallasite Eagle Station group, PMG = pallasite main group, R = Rumuruti-type, RI-rich = refractory inclusion-rich, T = tafassite \citep{Bouvieretal2021}, TIS = tissemouminite \citep{Stephantetal2023}, URE = ureilite, WIN = winonaite. "Metachondrites" represent type 7 primitive achondrites and chondritic impact melt rocks. Modified after \cite{Jacquet2022}.}\label{binominal}
\end{figure}

  Present-day chondrites do not show a continuum in compositional space, but discrete clusters known as \textit{chemical groups} \citep[e.g.][]{ScottKrot2014}. Chondrite groups are given letter designations, such as the H, L, and LL ordinary chondrite groups (originally meaning "high iron", "low iron", "low iron low metal"), which can be combined with a petrologic type in compact classifications such as "H5". Chemical groups are generally interpreted as representatives of distinct primary parent bodies (that is, prior to any collisional disruption; \cite{Greenwoodetal2020}), as sometimes supported by peaks in cosmic ray exposure ages suggesting common ejection events \citep{HerzogCaffee2014}. The lack of a compositional continuum is not simply a matter of incomplete sampling of the main belt asteroids by meteorites. In fact, the intermediate compositions, which likely existed in the protoplanetary disk, must have mostly disappeared, e.g. by incorporation into planets, since the present-day main belt, not exceeding a twentieth of lunar mass, is but a very partial sampling of the original population of planetesimals \citep{Jacquet2022}. Taking into account ungrouped meteorites (as the recognition of a group requires a minimum of five unpaired members; \cite{Wasson1974}), meteorites may represent 95-148 primary parent bodies \citep{Greenwoodetal2020}. Yet the distinction between primary parent bodies of origin may be difficult if some happened to accrete close-by in the disk and thus in compositional space, so the empirical concept of groups needs not be given a parent body interpretation \citep{Jacquet2022}, although it certainly has genetic significance at least in pointing to a restricted space-time section of origin in the disk. Groups may be further combined in higher-level taxa, sometimes called "clans", such as the ordinary chondrites, encompassing more inclusive regions of the disk ("reservoirs" in cosmochemical parlance). The current taxonomy is depicted in Fig. \ref{binominal}.

  At the highest level, there is a clear isotopic dichotomy between carbonaceous (C) and non-carbonaceous (NC) chondrites \citep{Warren2011b}. C chondrites are richer in many neutron-rich isotopes (but also in oxygen-16) than NC meteorites which are closer to terrestrial values. NC chondrites chiefly comprise ordinary chondrites (chiefly composed of olivine, pyroxene, iron-nickel metal and iron sulfides), and the very reduced enstatite chondrites (dominated by magnesian pyroxene (enstatite), iron-nickel metal and multifarious sulfides), which are isotopically closest to the Earth. Samples returned from S type asteroid Itokawa, which resembled LL4-6 chondrites \cite{Nakamuraetal2011}, confirmed the genetic link between ordinary chondrites and S type asteroids, with spectral signatures of Fe-bearing silicates olivine and pyroxene, which dominate inside 2.5 au in the main belt  \citep{DeMeoCarry2014}, the region primarily sampled by Earth-crossing meteoroids \citep{Binzeletal2015}.

  Carbonaceous chondrites (where group names always begin by "C", e.g. CI, CV, CM...), are spectroscopically similar to the low-albedo, presumably carbon-rich C complex asteroids, as confirmed by the recent return of CI-like samples from C asteroid Ryugu \citep{Nakamuraetal2023} and B asteroid Bennu \citep{Laurettaetal2024}. The prevalence of aqueous alteration effects among C chondrites may result from their accretion beyond the water snow line \citep{GrimmMcSween1989}. Micrometeoroids collected either in the stratosphere (and termed \textit{Interplanetary Dust Particles} or IDPs; \cite{Bradley2014}) or on the ground, in particular in Antarctic snow (\textit{micrometeorites}; \cite{Gengeetal2020}), do not always show evidence of aqueous alteration but are on the whole most comparable to carbonaceous chondrites, and are often even more carbon-rich \citep{Engrandetal2023}. Some may derive from comets, which may form a compositional continuum with C complex asteroids \citep{Gounelleetal2008}. Yet differences exist. Known comets are isotopically heavier in H and N, for example, than most carbonaceous chondrites \citep[e.g.][]{Aleon2010}, even though those of the CRHB clan, a subset of carbonaceous chondrites (often metal-rich, but refractory inclusion-depleted) show some isotopic affinities. So the C superclan has some structure. The CI may make another division, drastically underrepresented in the meteoritical record because of their friability judging from the Ryugu and Bennu sample returns, but it may be premature though to elevate them on a par with NC and other C in some disk-scale "trichotomy" \citep{Hoppetal2022}, as they may have contributed to both CRHB and "normal" refractory inclusion-rich carbonaceous chondrites ("CC-RI";\cite{Marrocchietal2022}), and are not more remarkable than the CRHB/CC-RI divide.

  Overall then, NC chondrites may originate from the inner disk and C chondrites from the outer disk (where the boundary, mobile or not, may have been at a few au (astronomical units), perhaps near Jupiter). In isotopic space, differentiated meteorites also plot in either the C or NC superclan \citep{Warren2011b,Kleineetal2020}, suggesting that their protoliths, whose isotopic composition would not change upon differentiation, resembled present-day C or NC chondrites%, and hinting at an early formation of the NC/C hiatus or the later removal of intermediate materials; \cite{Jacquetetal2019}
. Most differentiated meteorites belong to the NC superclan (if at more $^{16}$O-rich compositions than most NC chondrites) in line with more rapid accretion (and thus higher $^{26}$Al abundances) at shorter heliocentric distances. Most iron meteorite groups %(or grouplets not yet meeting the five-member threshold for formal recognition)
 do not seem to have achondrite, let alone mantle, counterparts \citep{Scottetal2015}. This suggests early stripping and loss of stony layers, perhaps in inner disk regions more collisional than the environment in which Vesta, the presumed parent body of howardites-eucrites-diogenites (HED), retained its largely intact basaltic crust.

\subsection{Chondrite components}

\begin{figure}[htbp]%
\centering
\includegraphics[width=\textwidth]{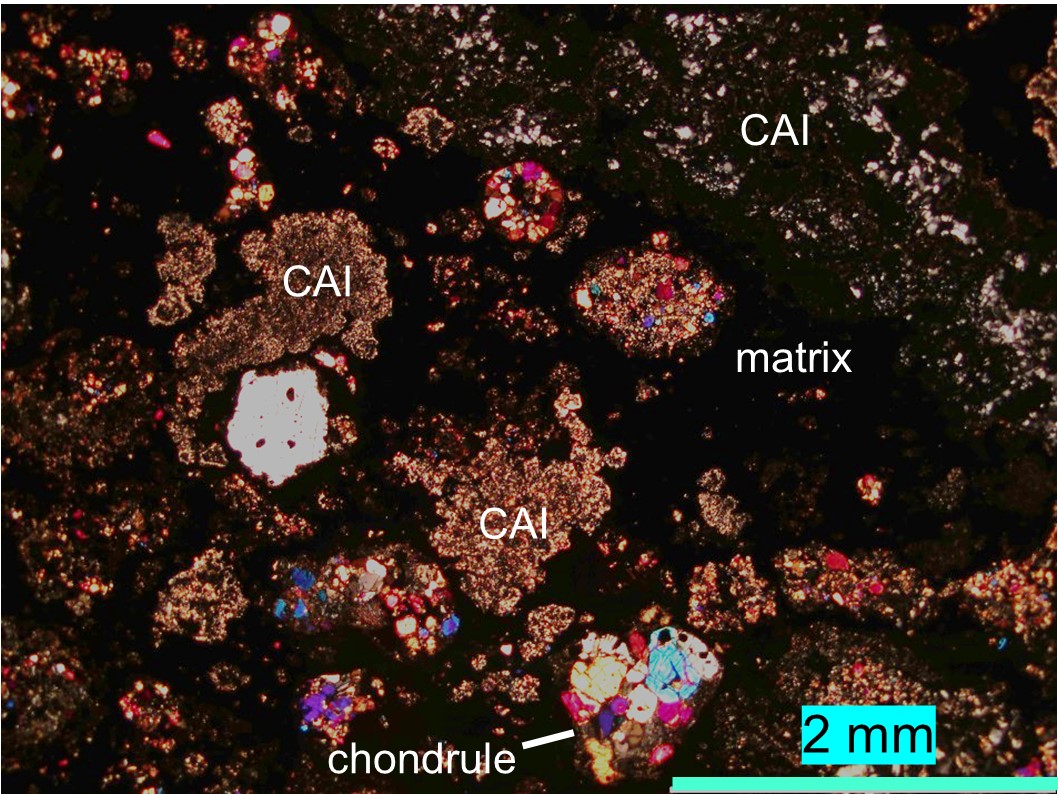}
\caption{Transmitted polarized and analyzed light microscopy of the Allende (CV3) carbonaceous chondrites. The polarization/analysis of the light yields interference colors which contribute to the identification of minerals. The CAI on the upper right mainly consists of coarse-grained melilite and has presumably undergone at least partial melting (unlike the two other fine-grained CAIs indicated). The matrix appears opaque because it is finer-grained than the thickness ($\sim$30 $\mu$m) of the thin section (3181lm2 of the Mus\'{e}um national d'Histoire naturelle in Paris).}
\label{Allende}
\end{figure}

\begin{figure}[htbp]%
\centering
\includegraphics[width=\textwidth]{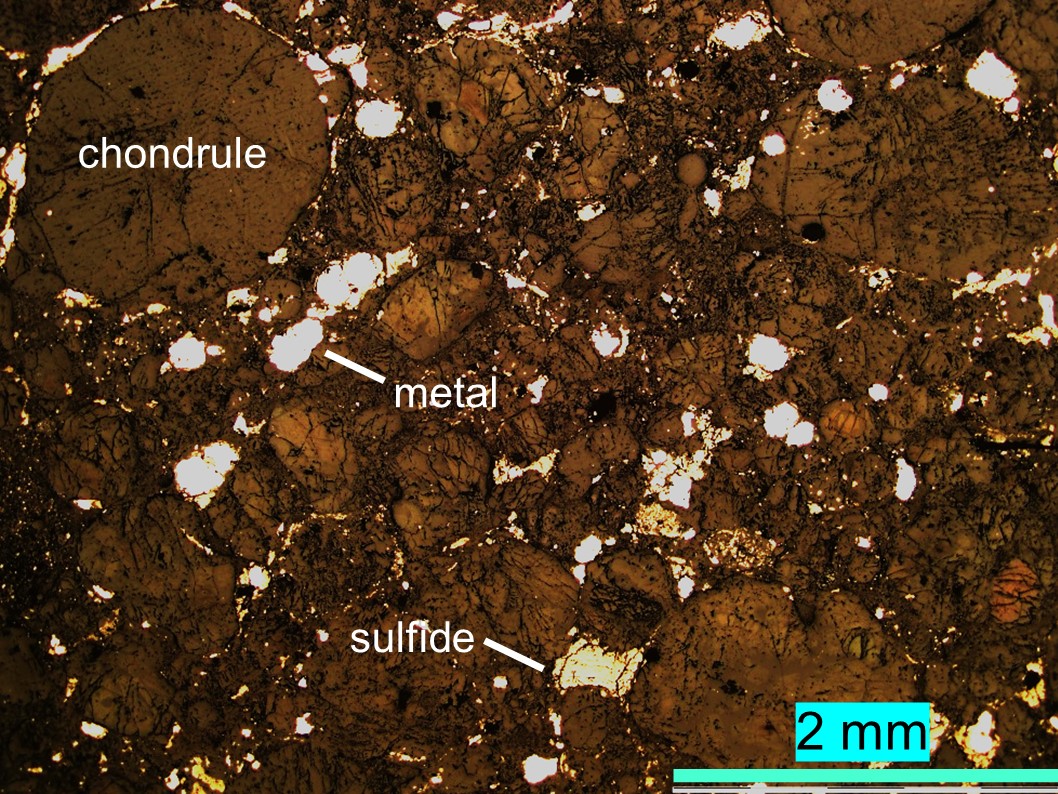}
\caption{Reflected light microscopy of the Hallingeberg (L3.4) ordinary chondrite (thin section 3576lm1 of the Mus\'{e}um national d'Histoire naturelle in Paris).}
\label{Hallingeberg}
\end{figure}

  Thus far, we have been mainly interested in the bulk composition of chondrites, but their individual components are the witnesses of protoplanetary disk processes prior to their accretion. Chondrites may indeed be described as "cosmic sediments", welding together varous millimeter or submillimeter-size components such as \textit{chondrules} and \textit{refractory inclusions} in a finer-grained \textit{matrix}. These are illustrated in the micrographs of two chondrites, carbonaceous and non-carbonaceous (Figures \ref{Allende} and \ref{Hallingeberg}) and will now be introduced individually.

\subsubsection{Refractory inclusions}

   Refractory inclusions are the oldest dated components formed in the Solar System ($\sim$4567-4568 Ma). The tight clustering of their initial $^{26}$Al/$^{27}$Al ratios around the canonical value $5.2\times 10^{-5}$ \citep[e.g.][]{Leeetal1976,Hussetal2001,SugiuraKrot2007,MacPhersonetal2012} suggests a formation epoch shorter than $\sim 10^5$ years. Refractory inclusions comprise the \textit{Calcium-Aluminum-rich Inclusions} (CAIs), which may be fine-grained aggregates of mineralogically zoned nodules or coarser-grained igneous objects, and the less refractory \textit{Amoeboid Olivine Aggregates} (AOAs) which are more or less annealed (forsteritic) olivine-rich and may contain CAI-like inclusions. Although some "lappige Chondren" (lobate "chondrules") were reported by \cite{Tschermak1885} in the Renazzo meteorite, the first extensive description of a CAI, which also made the connection with condensation models, was that of \cite{ChristopheMichelLevy1968} in the Vigarano meteorite. This was just one year before the massive fall of Allende, also a Vigarano-type carbonaceous chondrite (CV) laden with CAIs, definitively secured the attention of the meteoritical community for them \citep{Clarkeetal1971}. The CAI mineralogy corresponds indeed to the first phases expected to condense out of a cooling gas of solar composition (or the last residues to evaporate during heating), assuming thermodynamic equilibrium, with AOAs corresponding to the onset of the condensation of "main component" elements, in particular in the form of olivine and enstatite \citep{Hutchison2004,DavisRichter2014}. Mass-dependent light isotope enrichment in Si and Mg in fine-grained refractory inclusions is consistent with a kinetic condensation signature, perhaps within cooling timescales of weeks \citep{Marrocchietal2019AOA,Fukudaetal2021}, but many of them have undergone reheating, up to melting and partial evaporation \citep{MacPherson2005}. 

  Despite intergroup mineralogical variation, the common average isotopic signature of refractory inclusions in different chondrite groups (e.g. as $^{16}$O-rich as the solar wind; \cite{McKeeganetal2011}), indicate a common reservoir of origin. Although the high temperatures required for refractory inclusion condensation ($\gtrsim$1500 K) have long suggested an origin in the vicinity of the Sun, they are strikingly rare in non-carbonaceous chondrites ($\sim$0.01 vol\%; \cite{Dunhametal2023}) whereas they may make up to $\sim$10 vol\% of carbonaceous chondrites (in particular CVs; \cite{Ebeletal2016}), despite their presumed remoter region of provenance. Although C chondrites are isotopically intermediate between NC meteorites and refractory inclusions, mass balance calculations preclude making the former by simple addition of refractory inclusions to a NC-like endmember \citep{Nanneetal2019}. However, an isotopically CAI-like component with \textit{chondritic} chemistry could do \citep{Nanneetal2019,Schneideretal2020}; this may indicate a nonrefractory isotopically CAI-like complement mixed in the matrix and chondrule precursors.

\subsubsection{Chondrules}

    Chondrules make up $\sim$20-90 vol\% of chondrites (excepting the nearly chondrule-free CIs which have also undergone extensive aqueous alteration; \cite{Morinetal2022}), hence their names, ultimately derived from the Greek "khondros", "small grain". Chondrules are igneous spheroids, generally 0.1-1 mm in diameter, mostly made of the ferromagnesian silicates olivine and pyroxene set in a (more or less) glassy mesostasis \citep{Russelletal2018}. Opaque phases (that is, opaque in transmitted light microscopy) such as metal and sulfides also occur, and isolated opaque or silicate grains found in the interchondrule matrix may have been expelled from chondrules \citep{Zandaetal1994,Jacquetetal2013,Jacquet2021}. Most chondrules show porphyritic textures indicative of incomplete melting (with some identifiable relict grains from the precursor material) but the more spectacular nonporphyritic chondrules solidified from undercooled droplets. Ferromagnesian chondrules present a redox dichotomy between \textit{type I} chondrules, which are reduced (with very magnesian olivine and pyroxene and most Fe as metal) and more oxidized \textit{type II} chondrules (with Fe mostly in sulfides or silicates). Chondrule textures are consistent with formation by melting (incomplete for the most common porphyritic ones) and cooling within timescales variously estimated from minutes to days \citep{Jonesetal2018}. Al-Mg and Pb-Pb dating suggest chondrules are $\sim$1-3 Ma younger than CAIs \citep{KitaUshikubo2012}. Isotopic and mineralogical properties indicate chondrule provinciality in that chondrules from different groups were not produced as a single population, but were produced in several regions both in the C and NC reservoirs \citep{Jones2012, Schneideretal2020}. 

  Despite 222 years of research since their original description by Jacques-Louis de Bournon (in \cite{Howard1802}), the nature of the chondrule-forming mechanism(s) remains elusive. Proposed chondrule-forming scenarios are generally classified as either "nebular" or "planetary". While this is essentially a disciplinary division between astrophysics and geology, one may define these categories after the assumed nature of the chondrule precursors (that is, the material which was melted to produce chondrules), with nebular scenarios assuming free-floating mm-size aggregates while planetary scenarios derive them from already accreted planetesimals. Examples of nebular scenarios include shock waves passing through the disk \citep{Deschetal2005}, e.g. because of planetary embryos in eccentric orbits \citep{Boleyetal2013} or the expansion of an impact plume \citep{Stewartetal2024}, melting at the disk inner edge (X point; \cite{Shuetal1996}) or lightning discharges \citep{DeschCuzzi2000,JohansenOkuzumi2018} while planetary scenarios chiefly mean impact splashing \citep{Asphaugetal2011, SandersScott2012} or jetting \citep{Johnsonetal2015}, depending on whether the target material was already molten or not. Nebular scenarios are favored by evidence of refractory inclusion admixtures in chondrule bulk compositions \citep{MisawaNakamura1988,Schneideretal2020,Jacquet2022} and relict grains \citep{Krotetal2006,Marrocchietal2019relicts,Pirallaetal2021}. However, planetary scenarios fare better with evidence for large concentrations of condensable materials in chondrule-forming environments, at least two orders of magnitude above solar \citep{Tenneretal2015,CuzziAlexander2006,Jacquet2021,Alexanderetal2008}. A "planetary" (impact plume) origin is arguably consensual for the unusual (and mostly nonporphyritic) CB and many CH chondrite chondrules \citep[e.g.][]{Krotetal2005}, but this needs not apply to "mainsteam" chondrules. Thus, as yet, it is not known what significance chondrule formation had in the planet-forming processes, in particular whether it affected most planetesimal building blocks or only a late epiphenomenal minority \citep{HerbstGreenwood2019}. The \textbf{Marrocchi chapter} is devoted to this chondrule mystery.

\subsubsection{Matrix}

  The matrix of the most pristine chondrites comprises submicron-size grains. These include amorphous silicate phases, crystalline silicates and sulfides \citep{BrearleyJones1998}. Organic matter is also present, diffuse or as discrete globules, ranging from insoluble macromolecular, partly aromatic, material to soluble species such as aminoacids which have attracted interest regarding the genesis of life on Earth \citep{Gilmour2014}. Since matrix is most abundant in carbonaceous chondrites, their carbon content may be high (a few wt\%), hence their name historically, but this is not always the case. Water (or, strictly, hydroxyl ions) is also present in the amorphous silicates \citep{LeGuillouetal2015} or phyllosilicates and may result from melting of ice grains on the parent body (i.e. aqueous alteration), although preaccretionary hydration is possible \citep{Marrocchietal2023}. The porosity of CI chondrites, which are nearly pure matrix, reaches 34.9 vol\% \citep{Flynnetal2018} but matrix porosity may have been higher before lithification as a coherent rock, up to 70-80 vol\% judging from fabric in fine-grained rims around Allende chondrules \citep{Blandetal2011}.

  Some essentially monomineralic grains present very anomalous isotopic compositions, sometimes requiring log-scales in plots \citep{Zinner2014}. These grains, now termed \textit{presolar grains}, are thought to have condensed in the envelopes of dying stars (red giants, supernovae...) before the formation of the Sun, and as such they inherited the isotopic signatures of their characteristic nucleosynthetic processes. They predate CAIs (the oldest solids \textit{formed} in the Solar System) by millions to hundreds of millions of years according to cosmic ray exposure estimates \citep{Hecketal2020}. Although they now represent only a minority (less than a thousandth; \cite{Haenecouretal2018}) of chondrite matter, the original dust of the Solar System was evidently entirely presolar. Most likely, the isotopic variations for many elements among bulk meteorites are ultimately traceable to various proportions of different populations of presolar grains, whether by differential destruction during "thermal processing" \citep[e.g.][]{Niemeyer1988, VanKootenetal2016} or uneven original distribution in the parental molecular cloud and/or the disk \citep{Jacquetetal2019}. Specifically, C chondrites may have incorporated a greater proportion of grains enriched in r process isotopes (i.e. formed through \textit{r}apid neutron additions, in neutron-rich environments, e.g. supernovae) in contradistinction to s(low) process carriers (see \textbf{Liu chapter}). We note that bulk isotopic variations in oxygen do not seem to relate to varying nucleosynthetic contributions as originally proposed by \cite{Claytonetal1973}, but rather to one as yet elusive mass-independent process, such as CO self-shielding, which may have produced $^{16}$O-poor water exchanging isotopes with originally isotopically CAI-like nebular rocks \citep{Youngetal2008}.  

  We do not know what proportion of matrix (or IDPs) is still presolar \citep{Alexanderetal2017, Bradleyetal2022}, for some grains may be too small to be resolvable by NanoSIMS, and grains formed in the interstellar medium itself, as opposed to stellar atmospheres/ejecta, would not be readily identifiable isotopically (although this is suggested for D- or $^{15}$N-rich "hotspots" in organic matter; \cite{Gilmour2014}). Part of the matrix may relate to the condensation events which formed refractory inclusions, and another still may be related to chondrules, whether as small chondrule fragments or disequilibrium condensates. The contention that matrix derives from the same solar (in some respect) reservoir as chondrules and have exchanged elements with them in an overall closed system is termed chondrule-matrix complementarity and will be discussed in the \textbf{van Kooten chapter}. Matrix may form distinctive fine-grained rims %(FGRs)
 around chondrules, which some consider as dust accreted immediately after chondrule formation \citep{Zanettaetal2021,Pintoetal2022}%, although \cite{Metzleretal1992} proposed that interchondrule matrix in CM chondrites derived from FGR attrition on the parent body
.

\section{Dust in the early solar system}

\subsection{Early condensates in the solar System}

  What did the solar protoplanetary disk look like? One practical reference is the Minimum Mass Solar Nebula (MMSN; \cite{Hayashi1981}) where present-day planetary masses are mentally supplemented by solar gas to restore solar abundances. With some smoothing, one obtains a surface density $\Sigma_g$ profile \citep{Hayashi1981}:
\begin{equation}
\Sigma_g=1.7\times 10^4\:\mathrm{kg/m^2}\left(\frac{1\:\rm au}{r}\right)^{1.5}
\end{equation}
with $r$ the heliocentric distance. The MMSN would be about 0.01 M$_\odot$ in mass. This of course ignores any movement and loss of disk solids and gas to the Sun. In fact, simulations of disk building would predict an initial disk mass one order of magnitude higher \citep[e.g.][]{YangCiesla2012,Pignataleetal2018}.

  The MMSN is typically associated a temperature $T$ profile corresponding to passive heating by unattenuated sunlight\citep{Hayashi1981}:
\begin{equation}
T=280\:\mathrm{K}\left(\frac{1\:\rm au}{r}\right)^{0.5}
\end{equation}

This is however an overestimate, because of absorption of direct sunlight by the intervening dust would make the actual midplane temperature a factor of $\sim$2 lower at a few au \citep[e.g.][]{ChiangGoldreich1997}. Even then, only the inner boundary of the disk would be hot enough for refractory inclusion formation, as in the X wind model \citep{Shuetal1996}. However, the dissipation of turbulence can heat the disk to higher temperatures. For a steady disk of mass accretion rate $\dot{M}$, the corresponding contribution is \citep{Jacquetetal2012S}:
\begin{eqnarray}
T &=& \left(\frac{3}{128\pi^2}\frac{\kappa m}{\sigma k_B \alpha}\dot{M}^2\Omega_K^3\right)^{1/5}\nonumber\\
&=& 500\:\mathrm{K} \left(\frac{\kappa}{0.5\:\rm m^2/kg}\right)^{1/5}\left(\frac{10^{-3}}{\alpha}\right)^{1/5}\left(\frac{\dot{M}}{10^{-8}\rm M_\odot/a}\right)^{2/5}\left(\frac{1\:\rm au}{r}\right)^{9/10}
\end{eqnarray}
with $\kappa$ the (Rosseland mean) specific opacity, $k_B$ the Boltzmann constant, $m$ the mean molecular mass (2.3 amu for a solar gas), $\Omega_K=\sqrt{GM_{*}/r^3}$ the Keplerian angular velocity (with $G$ the gravitational constant and $M_{*}$ the stellar mass) and $\alpha$ the \citet{ShakuraSunyaev1973} turbulence parameter\footnote{The definition of $\alpha$ is more general than the case of turbulence (see e.g. \cite{Jacquet2013}) but the equations involved herein will assume this type of flow, with a Schmidt number (the ratio of kinematic effective viscosity to turbulent diffusivity) of unity for simplicity.}. Such "viscous heating" (termed such because, ideally, the large-scale effects of turbulence may be equated with that of an enhanced viscosity) must dominate at short heliocentric distances, and for early (Class 0 or I) disks (with high $\dot{M}$), may ensure CAI-forming temperature up to 1 au \citep[e.g.][]{YangCiesla2012,Pignataleetal2018,OwenJacquet2015}. In fact, evidence of rapid condensation of refractory inclusions \citep{Marrocchietal2019AOA} suggests that they formed in localized events under a lower ambient temperature, which thus may have been beyond their nominal condensation isotherms. 

  While refractory inclusion formation may thus have occurred over a wider extent of the early disk than previously believed, their presence at the large heliocentric distances (several au) housing the C superclan  still requires outward transport.  Disk winds have difficulties in lifting particles larger than a few tens of microns \citep[e.g.][]{Marrocchietal2019AOA,Deschetal2021} in a MMSN disk, but early disks in formation may have replenished the envelope with mm-size dust that way \citep{Tsukamotoetal2021}. Despite difficulties inherent to the survival of solids up to and from the disk inner edge \citep{Deschetal2010}, the X-wind model, or something in this vein, is still occasionnally referred to in the cosmochemical litterature as a way to leapfrog the (NC) inner disk but this could hardly be an expected effect given the sensitivity of the transport range to size \citep{Hu2010}. Turbulent diffusion alone may be insufficient \citep[e.g.][]{Ciesla2010}, but if the solar protoplanetary disk started off sufficiently compact, a viscous spreading could have incurred outward mean velocities in the vicinity of the CAI-forming region \citep{Jacquetetal2011CAI,YangCiesla2012,Pignataleetal2018,MarschallMorbidelli2023}.

\subsection{Dust growth}
Whether condensed in the disk or inherited from the parental molecular cloud, the dust first grows by coagulation, which is a very
complex process \citep{Birnstieletal2016}. Small grains stick to each other by van der Waals forces. As a rule of thumb the
time scale of growth is \citep[cf.][]{Birnstieletal2012}
\begin{equation}
\tau_{\mathrm{grow}} \equiv\frac{a}{\dot a} \sim \frac{1}{\Omega_K\epsilon}
\end{equation}
where $a$ is the radius of the dust grain, $\dot a$ its time derivative,
 and $\epsilon$ the dust-to-gas mass ratio, about $0.01$ for a solar gas. Given that the growth time scale $\tau_{\mathrm{grow}}$ does not depend on the grain size, the growth process (at least in this simplification) is exponential. 
%The total time of exponential growth $t_{\mathrm{grow}}$ needed to grow from size $a_{\mathrm{min}}$ to $a_{\mathrm{max}}$ is then
%\begin{equation}
%\frac{t_{\mathrm{grow}}(a_{\mathrm{min}},a_{\mathrm{max}})}{P} = \frac{1}{2\pi\epsilon}\ln\left(\frac{a_{\mathrm{max}}}{a_{\mathrm{min}}}\right)
%\end{equation}
%where $P=2\pi/\Omega_K$ is the orbital period. 
 The growth from micron size to millimeter size, for $\epsilon=0.01$, thus takes about a hundred orbits. Around this size, however, collision velocities, due to turbulence or differential drift \citep{Birnstieletal2012}, are too high for sticking, and may rather lead to bouncing, erosion or fragmentation, so further growth by coagulation alone is frustrated \citep[e.g.][]{Blum2018}. 

  The typical size of chondrules or refractory inclusions may thus be accounted by this coagulation bottleneck \citep{Jacquet2014size}. In principle, fluffy aggregates of submicron monomers might grow fractally to planetesimal masses and beyond \citep{Okuzumietal2012}, but the sticking behaviour beyond cm size is untested experimentally \citep{Simonetal2022}. The very existence of refractory inclusions and chondrules also implies that chondrite parent bodies did not simply grow from such ultraporous aggregates, but have always had nonnegligible filling factors ($\gtrsim 0.1$; e.g. \cite{Blandetal2011}), but the question may be left open for cometesimals.

\subsection{Drift}

  In the same way snowflakes drift downward (toward higher pressures) relative to a wind carrying them, once dust aggregates reach millimeter size, their drift becomes significant. Their radial velocity is \citep{Birnstieletal2012}:
\begin{equation}
v_{r,d} %= \left(\frac{d\ln p}{d\ln r}\right)\frac{c_s^2}{v_K}\, \mathrm{St}
=\left(u_r+\frac{t_f}{\rho_g}\frac{\partial p}{\partial r}\right)\left(1+\mathrm{St}^2\right)^{-1}    %\left((1+\epsilon)^2+\mathrm{St}^2\right)^{-1}
\end{equation}
where $p=\rho_gc_s^2$ is the midplane gas pressure, $\rho_g$ the gas density, $u_r$ the gas radial velocity and $c_s=\sqrt{k_BT/m}$ the isothermal sound
speed of the gas. %The symbol 
 Finally,
$\mathrm{St}\equiv\Omega_Kt_{f}$ is the Stokes number of the dust particle,
 with $t_f$ the friction time scale, also often called the
stopping time.  %Particles with $\mathrm{St}<1$ have a stopping time shorter than the orbital time and tend to settle to the midplane. The smaller the $\mathrm{St}$, the slower the settling speed. Particles with $\mathrm{St}>1$ have a stopping time longer than the orbital time. %, and they tend to perform a damped orbital oscillation about the midplane.
%When one speaks of ``pebbles'', or generally when one discusses dust coagulation, one can generally assume that $\mathrm{St<1}$. 
The limit $\mathrm{St}\gg1$ corresponds to planetesimals, which essentially follow Keplerian motion. %have their own orbital elements and obey N-body dynamics. 

  For particles smaller than the gas molecular mean free path, as appropriate for mm-size objects, the Epstein drag law applies and, at the midplane
\citep{Birnstieletal2010}):
\begin{equation}
\label{St}
\mathrm{St} = \frac{\pi}{2}\frac{\rho_{\bullet}}{\Sigma_g}\, a
\end{equation}
where $\rho_{\bullet}$ is the effective material density of the dust aggregate,
with porosity accounted for%, and $\Sigma_g$ the surface density of the gas
. The Stokes number is therefore a dimensionless
measure of the grain radius $a$.
  For St$\ll$ 1 and negligible $u_r$, we have:

\begin{equation}\label{eq-vrd-formula}
v_{r,d} \approx \frac{\partial\ln p}{\partial\ln r}\frac{c_s^2}{v_K}\, \mathrm{St}=-0.1 \mathrm{m/s} \left(\frac{a}{\rm 1\: mm}\right)\left(\frac{\rho_{\bullet}}{2\rm \:g/cm^3}\right)\left(\frac{r}{\rm 3\: au}\right)^{1.5}
\end{equation}
with $v_K=\sqrt{GM_{*}/r}$ the Kepler velocity. The numerical evaluation assumes the MMSN. % (MMSN; \citet{Hayashi1981}). %The double-logarithmic derivative in Eq.~(\ref{eq-vrd-formula}) depends on the radial structure of the protoplanetary disk. For a standard powerlaw disk model with gas column density $\Sigma_g$ and temperature $T$ going as
%\begin{equation}
%\Sigma_g\propto r^{-p},\qquad T\propto r^{-q}
%\end{equation}
%one can derive that the midplane gas {\em volume-}density $\rho_g$ goes as $\rho_g\propto r^{-p-(3-q)/2}$, where we used that the pressure scale height of the disk is $h_p=c_s/\Omega_K$.  The gas pressure at the midplane therefore goes as $p\propto r^{-p-(3+q)/2}$. Consequently,
%\begin{equation}
%\left(\frac{d\ln p}{d\ln r}\right) = -p-\frac{3+q}{2}
%\end{equation}
%For the minimum mass solar nebula (MMSN) with $p=3/2$ and $q=1/2$, this becomes $d\ln p/d\ln r=-3.25$, while for a standard viscous disk with $p=1$ and $q=1/2$, this becomes $d\ln p/d\ln r=-2.75$.For the MMSN with $\Sigma_g\simeq 330\,$g/cm$^2$ and $T\simeq 170\,$K at $r=3\;\mathrm{au}$, a dust aggregate with $a=0.1\,$cm and $\rho_{\bullet}=2\,$g/cm$^3$ will have a Stokes number of $\mathrm{St}\simeq 10^{-3}$. From Eq.~(\ref{eq-vrd-formula}) one thus finds that the radial drift velocity at $r=3\;\mathrm{au}$ is about -11 cm/s. 
The radial drift time scale
%$\tau_{\mathrm{drift}}=r/|v_{r,d}|$
 of a 1 mm dust particle at
$r=3\;\mathrm{au}$ is thus about $10^5$ years.

This short time scale shows that radial drift is a major problem in the
planet-forming region of the protoplanetary disk. It has long been known from telescopic observations that mm-size "pebbles" do survive in the outer regions ($r\gg 30$ au) of
disks of millions of years of age. Within our own solar system, single chondrites can host chondrules and refractory inclusions with age differences much larger than 0.1 Ma
\citep{Villeneuveetal2009,Connellyetal2017}. This suggests that the material that eventually forms an
asteroid must have been able to withstand the radial drift for prolonged
stretches of time.

\subsection{Dust trapping}

\begin{figure}
  \centerline{\includegraphics[width=0.49\textwidth]{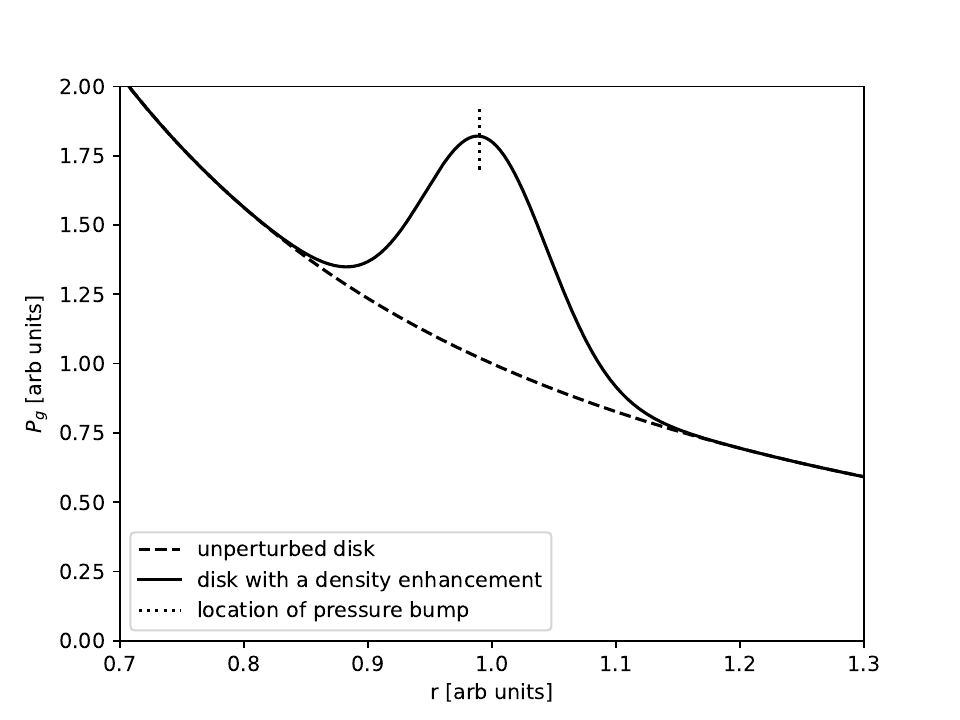}
  \includegraphics[width=0.49\textwidth]{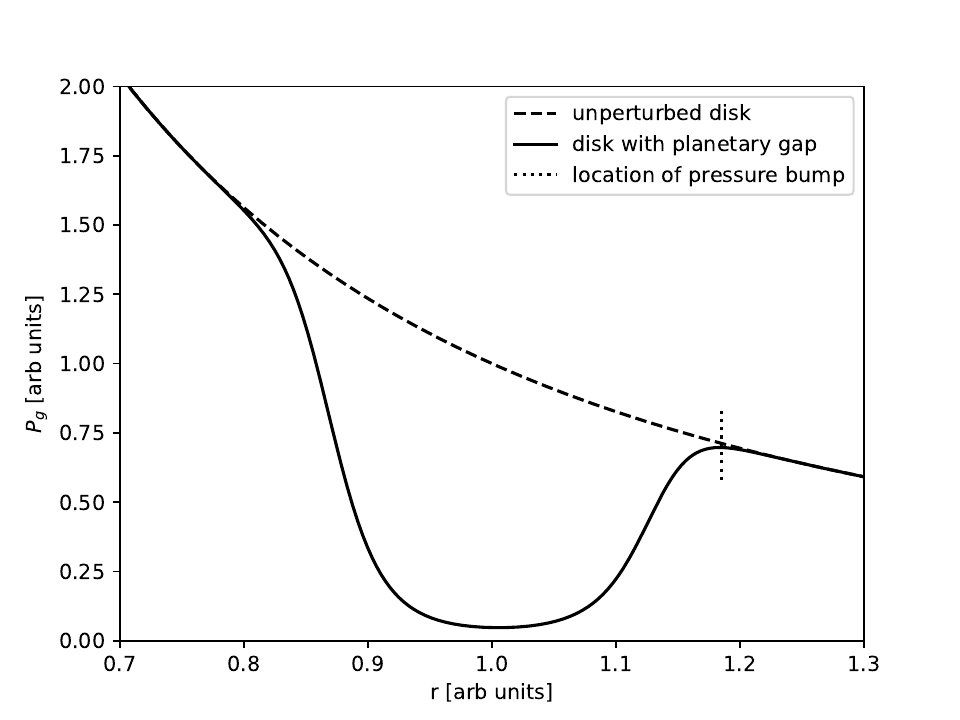}}
  \caption{\label{fig-bump-types}Sketches of how an annular density
    enhancement (left) or a planetary gap (right) can lead to a local
    pressure maximum (`pressure bump'), which can trap dust.}
\end{figure}

One solution is the presence of one or more local maxima in the radial gas pressure profile (Fig.~\ref{fig-bump-types}). At the peak of one such local maximum, one has
$\partial\ln p/\partial\ln r=0$ (no drift), and inside of this peak one has even $\partial\ln p/\partial\ln
r>0$ (outward drift). Such a local ``pressure bump'', which manifests itself
generally as a ring in a Keplerian disk, tends to accumulate dust
particles. They drift toward the peak from both sides, and get trapped
there \citep{Pinillaetal2012}. This allows these particles to stay there indefinitely, as long as the
local pressure bump is long-lived.

\begin{figure}
  \centerline{\includegraphics[width=\textwidth]{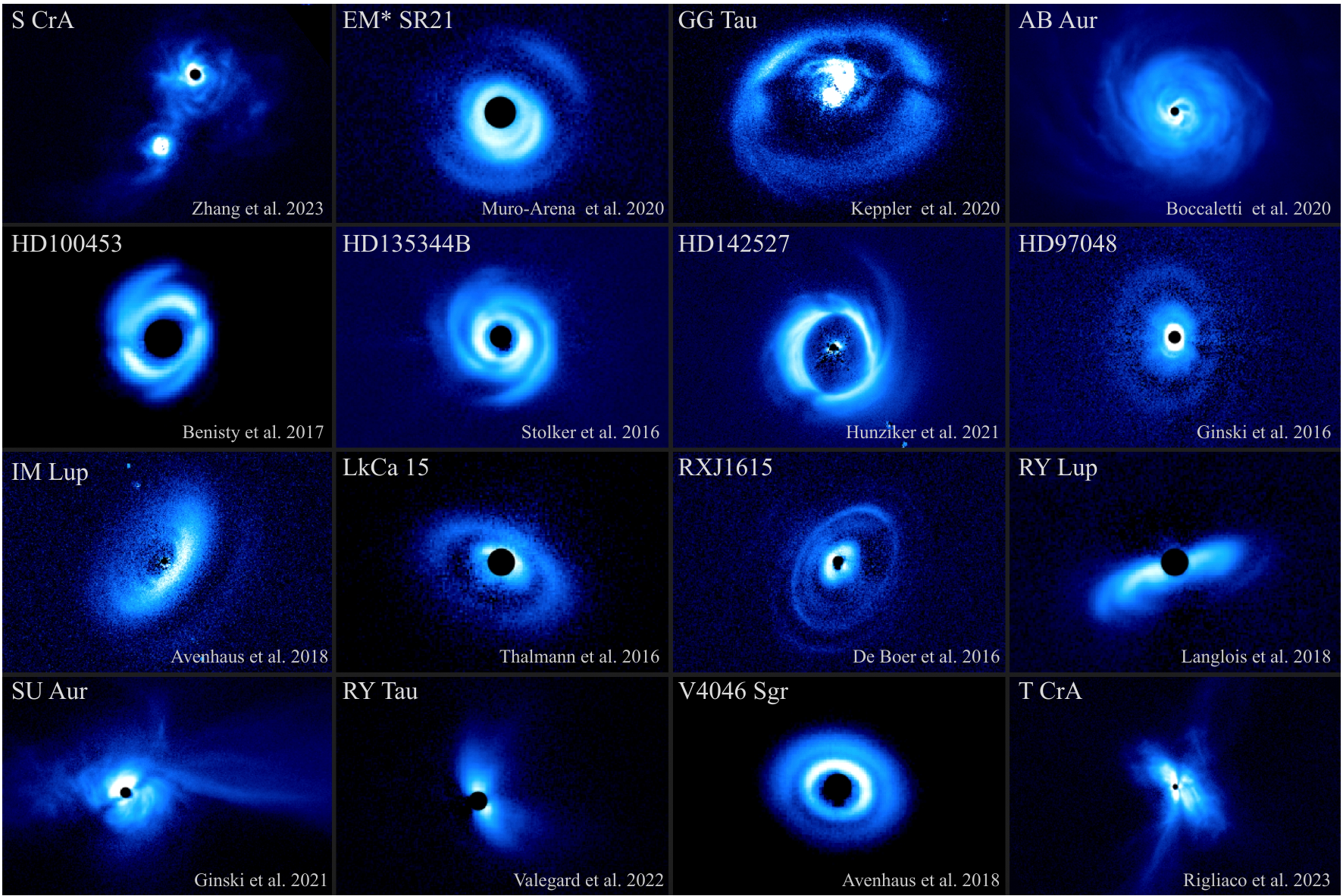}}%sphere_images.pdf}}
  \caption{\label{disk-images-vlt-sphere}Protoplanetary disk images taken with
    the SPHERE instrument on the VLT at optical wavelengths. Image credit: C. Ginski.}
\end{figure}

\begin{figure}
  \centerline{\includegraphics[width=\textwidth]{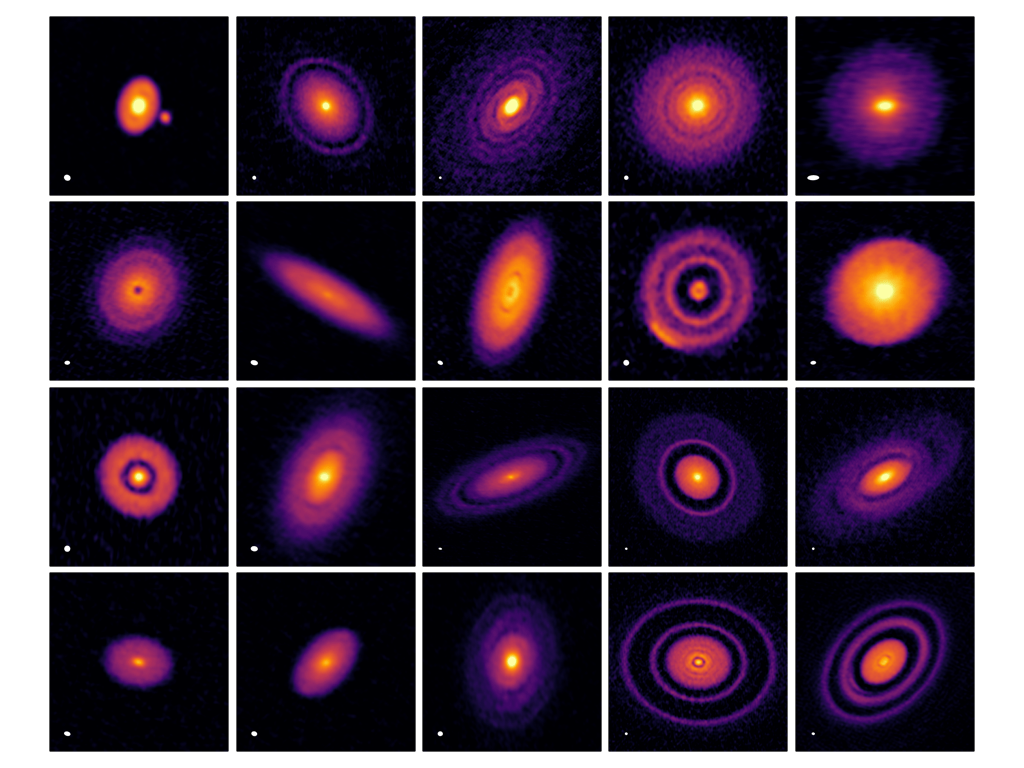}}
  \caption{\label{disk-images-alma}Protoplanetary disk images taken with ALMA at
    millimeter wavelength. The images shown here are from the DSHARP campaign
    \citep{Andrewsetal2018}.}
\end{figure}

  Indeed, observations suggest smooth disks are the exception rather than the rule. VLT images
 of protoplanetary disks around young stars about 140 parsec away \citep[][and
  Fig.~\ref{disk-images-vlt-sphere}]{Benistyetal2023} reveal a
rich variety of substructures in these disks. Many disks display several
concentric rings in their surface layers, while others show complex spiral
patterns. Some of these disks show evidence of being warped, i.e.\ non-planar,
and some show signatures of ongoing infall of matter from farther away
\citep[e.g.,][]{Ginskietal2021}. The ALMA images (Fig.~\ref{disk-images-alma}) show that the large mm-size dust grains are arranged mostly in
concentric rings. These rings are most likely ``dust traps'', in which the dust
grains are trapped at circular local pressure maxima
\citep{Dullemondetal2018}. Indeed, the larger the dust grain, the more it drifts toward pressure maxima.

  Possibly then, such dust traps may have isolated several reservoirs in the solar nebula, in particular the NC and C reservoirs. The CAI enrichment in the C reservoir could be due to such trapping, while most inner disk CAIs would have been lost to the protosun \citep{Kruijeretal2020,Deschetal2018}. However, the similar level of Mo isotopic anomalies shown by NC irons and chondrites suggest that CAIs (or at least CAIs isotopically similar to their CC counterparts) cannot have been abundant early on there, perhaps as a result of their removal by an early outward flow \citep{Jacquetetal2011CAI, MarschallMorbidelli2023}.
 %An earlier accretion of C chondrites before CAI drift \citep{Jacquetetal2012S} seems at variance with dating of chondrules and chondrites \citep{KitaUshikubo2012,SugiuraFujiya2014}.
 A pressure bump would still allow fine dust (tightly coupled to the gas) to go through \citep{Haugbolleetal2019,Stammleretal2023}, in the direction of the flow relative to it, so some transfer of material would remain possible, and may possibly account for an apparent secular isotopic evolution of the NC superclan in the direction of C meteorites \citep[e.g.][]{Schilleretal2018}, though the putative apex would lie outside the range of sampled C meteorites \citep{Burkhardtetal2021}. In the C superclan, an evolution from chondrule- and refractory inclusion-rich CCs toward inclusion-poor CI may have been incurred by differential drift \citep{Hellmannetal2023}.  Within each dust trap, so long it is open to influx of drifting or infalling dust, isotopic variation might then be essentially a matter of accretion time. Unfortunately, ALMA does not have
the spatial resolution to see whether dust rings occur at distances between 1 and 5 au from a young star.

  The cause of such local ring-shaped pressure maxima in a
protoplanetary disk remains a topic of debate. The most natural explanation
could be the existence of a planet \citep{Dipierroetal2015, Dongetal2015}. If a planet is massive enough to open a gap,
it will produce a local pressure maximum at the outer boundary of the gap (Fig.~\ref{fig-bump-types}-right). This
will halt the radial drift of millimeter size dust aggregates. In planet
formation theories, the minimum planetary mass to produce such a dust trap is
called the ``pebble isolation mass''
\citep{Lambrechtsetal2014,Bitschetal2018}, which can be roughly written as
\begin{equation}
M_{\mathrm{iso}} = 20\,\left(\frac{h_p/r}{0.05}\right)^3\,M_{\oplus}
\end{equation}
where $M_{\oplus}$ is the Earth mass and $h_p=c_s/\Omega_K$ the  pressure scale height of the disk. For the MMSN, $h_p/r$ is between 0.03 (near 1 au) and 0.06 (near 10 au). This
implies that as Jupiter was still forming, it would already early on (at less
than 10\% of its final mass) produce a pressure bump just outside of its orbit
that stops radial drift of pebbles. As Jupiter grew more massive, it might even have caused additional pressure bumps inward of its orbit, thus
trapping dust there as well \citep{Baeetal2017}. However, whether this
multi-trap phenomenon is viable depends on the details of the thermodynamics of
the disk \citep{Ziamprasetal2020}.

There are, however, still some problems with the planetary origin of dust
traps. While massive planets certainly produce dust traps, not all dust traps
can be explained well with planets. For instance, %as was noted during the workshop by A.~Morbidelli, 
Uranus and Neptune, according to the above formula,
have masses well below the pebble isolation mass, and cannot thus produce a
dust trap. This means that they cannot have prevented radial drift of material
that produced the Trans Neptunian Objects. Also, since dust particles grow
within about a hundred orbits (see above), radial drift may already set in
before the first planets have grown massive enough to stop it.

  There are therefore numerous alternative planet-free models being suggested that could lead
to local pressure bumps. Should infall proceed along streamers e.g., pressure bumps may be created where they land on the disk \citep{Kuznetsovaetal2022}, but these may diffuse away once the streamers cease. Radial gas transport mechanisms inside the disk, if of unequal efficiency as a function of heliocentric distance, may also lead to "traffic jams".

  Radial transport close to the Sun is generally attributed to turbulence driven by the magnetorotational instability (MRI). If the turbulent viscosity is a sufficiently strong function of the magnetic pressure, weakly magnetized zones will accumulate matter at the expense of the more magnetized ones \citep[e.g.][]{Iwasakietal2024}. The MRI, however, cannot operate at all distances, for the required ionization levels are not ensured by thermal ionization below about $\sim$1000 K, so the MRI-active inner region must give way to a MRI-free "dead zone" overlain by active layers ionized by nonthermal radiations from the Sun or cosmic rays \citep{Gammie1996}. Since the effective viscosity drops, a pressure bump may arise at the dead zone inner edge \citep{Flocketal2015}.

  A sublimation line, such as the water ice line, could also incur short-scale variations in the effective viscosity. The viscosity $\nu=\alpha k_BT/(m\Omega_K)$ may drop inside because of the increase in molecular mass $m$ (assuming a constant $\alpha$; \cite{Charnozetal2021}). Outside, it may also increase because the temperature $T$ will be enhanced by condensing grains increasing the local opacity \citep{CieslaCuzzi2006}, but those grains may also lower the effective $\alpha$ by scavenging electrons available in the active layers \citep{KretkeLin2007}. However, transport in the dead zone may be driven by magnetized winds leaving the disk rather than disk turbulence. Those winds may also produce large-scale ring-like pressure bumps  \citep{Riolsetal2020, Surianoetal2018}. Since the simulations do not extend to more than a few hundred orbits, it is unknown whether such bumps can be long-lived.

\section{Planet(esimal) formation}
%Formation of the gravitationally bound bodies from which the contemporary asteroids and comets derive is perhaps the most elusive aspect of planet formation theory. There are two major pathways considered in the literature: bottom-up growth of ever larger dust aggregates until kilometer sizes and gravitational collapse of overdense dust clouds leading directly to planetesimals. Currently, the latter pathway seems to be favored in the literature, in particular thanks to the numerous studies of the so-called streaming instability, which we focus on here. A recent comprehensive review of the planetesimal forming mechanisms is provided by \citet{Simonetal2022}.%2022arXiv221204509S}.

\subsection{Dust concentration}
Stopping the drift of dust is not enough to accrete planetesimals. We saw earlier that simple growth by sticking reached a bottleneck around mm-cm size. The mutual gravity of dust grains would be helpful, but to overcome 
%As discussed in Sect.~{\bf{1.2 from Kees' part}}, in the protosolar disk, the dust-to-gas ratio was about 1\%. For the planetesimal formation to be possible, dust must have been concentrated to the point where its gravitational attraction overcame 
competing effects such as shear of the differentially rotating disk% and turbulent diffusion
, one needs to reach a critical density called the Roche density, defined, for a fluid, as \citep{Chandrasekhar1967}
\begin{equation}\label{Rochedens}
    \rho_{\rm{R}} = 3.5 \frac{M_\star}{r^3} = 2\cdot10^{-6} \frac{M_\star}{M_{\odot}} \left(\frac{r}{\mathrm{au}}\right)^{-3}\ \mathrm{g/cm^3},
\end{equation}
(The exact threshold applicable in a dusty disk is somewhat controversial, and could be a factor of 5 lower for an axisymmetric ring of tightly coupled dust and gas, see e.g. \citet{ChiangYoudin2010}). In the MMSN, the Roche density translates into a midplane dust-to-gas ratio of
\begin{equation}
    \epsilon = 1530\cdot\left(\frac{r}{\mathrm{au}}\right)^{-1/4},
\end{equation}
five orders of magnitude above the solar value. To reach such a high concentration, we may count on dust settling toward the midplane, since pressure is maximum there, for a given radial location. For settling to be significant, dust must grow to larger aggregates with $S\equiv\mathrm{St}/\alpha>1$ \citep{Jacquetetal2012S}. Indeed, the competition between settling and turbulence results in a midplane density enhancement factor of $(1+S)^{1/2}$. In astrophysical literature (and quite irrespective of geological usage), such aggregates are called "pebbles", with a Stokes number (see equation \ref{St}) falling in between 0.001 and 1. The corresponding size range depends on gas density but at the distance of the asteroid belt, pebbles are typically at least millimeter-sizes and this minimum size increases as we move closer to the star or consider a disk more massive than the MMSN. Thus, it is unclear whether individual chondrules can be such pebbles or not. Most likely, aggregates of chondrules and matrix dust may be needed to reach the pebble sizes. %Possibly, the formation of chondrules, more robust than dust aggregates in collisions, may have fostered further growth. 
In this case, an early coagulation of the latter may be a way to preserve chemical complementarity until parent body accretion \citep{Jacquet2014size}, but motion of independent chondrules and dust grains would not have destroyed complementarity if the $S$ of the chondrules was $<1$ \citep{Goldbergetal2015}, so matrix-chondrule complementarity might not be a stringent constraint on the dynamics of chondrite components. 

  %Whatever their exact nature, pebbles efficiently settle to the midplane of the disk, which alone increases their density. %The enhancement factor at the midplane can be estimated as $(1+S)^{1/2}$. 

  For typical values of ${\rm{St}}=10^{-2}$ and $\alpha=10^{-4}$, settling causes the dust-to-gas ratio to increase by one order of magnitude and Kelvin-Helmholtz instabilities may then prevent further settling \citep{Weidenschilling1980}. However, further enhancement is needed to proceed with planetesimal formation, which must be in radial and/or azimuthal direction.  
 There are a range of processes discussed in the literature that lead to such enhancements. We have already mentioned axisymmetric pressure bumps for radial pileup. Vortices invoked for explaining non-axisymmetric features in observed disk may concentrate dust in azimuthal direction \citep{Raettigetal2015,Lietal2020}. In a turbulent cascade, dust aggregates may get concentrated by the gas eddies \citep{HartlepCuzzi2020}. However, it is unlikely that any of these processes alone leads to dust concentration up to the Roche density. 

\subsection{Streaming instability}

\begin{figure}[t]
  \centerline{\includegraphics[width=\textwidth]{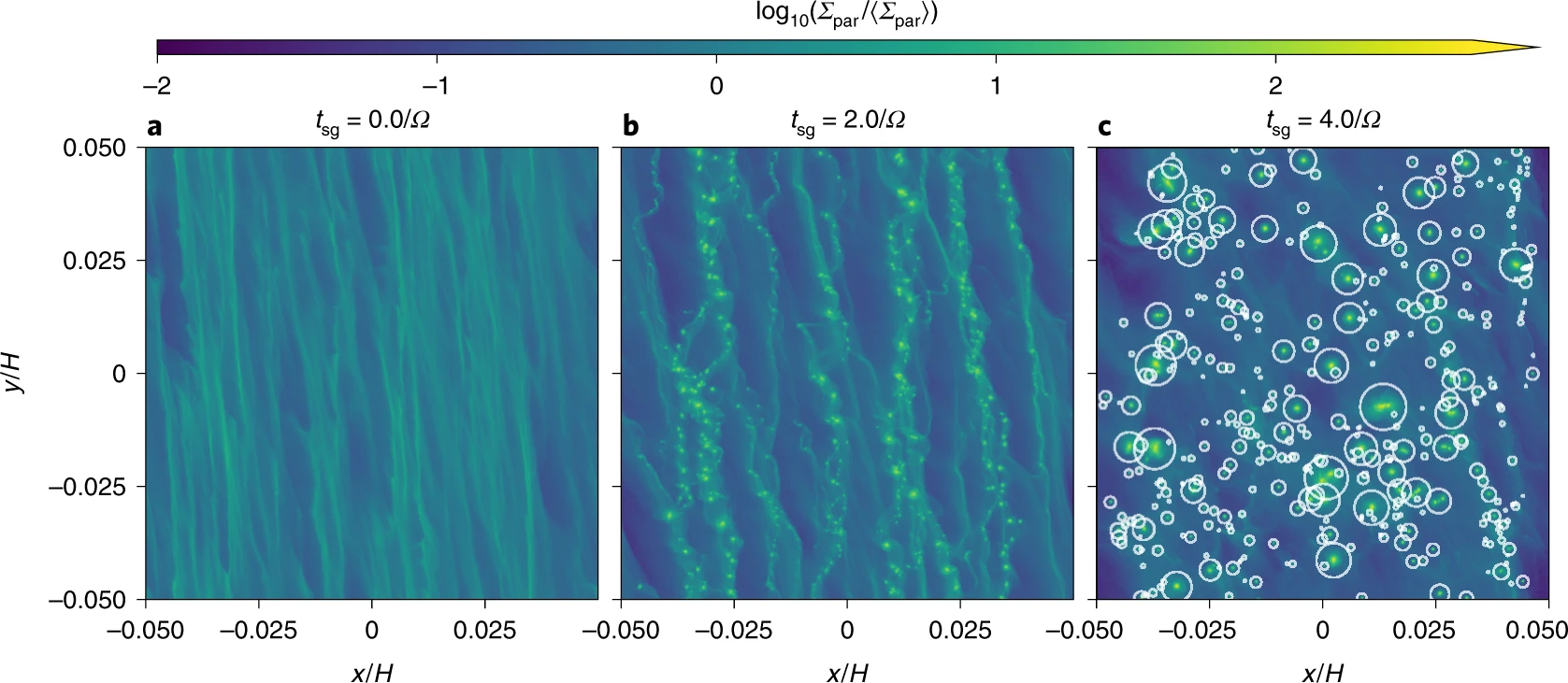}}
  \caption{\label{fig:SI} Streaming instability simulation from \citet{Nesvornyetal2019}%NatAs...3..808N}
. The panels show the time evolution, in the units of local orbital time counted after particle self-gravity was switched on, of the vertically integrated density of pebbles relative to the initially uniform surface density. The $x$ and $y$ coordinates show the dimensions of the simulated domain in units of the gas disk scale height. The filaments already present at the time of switching on self-gravity fragment into pebble clumps. The circles in the last panel show the Hill spheres of gravitationally-bound clumps.}
\end{figure}

  Now, if only a solid/gas ratio of order unity (i.e. two orders of magnitude above solar) could be achieved, the \textit{streaming instability} could take over to bring the pebble clumps to the Roche density, depending on the value of the Stokes number. The streaming instability is the instability of the dust drift in the sub-Keplerian gas disk %({\bf{see Sect.~1.2 from Kees' part}})
 to local overdensities when the momentum feedback from dust to gas is taken into account \citep{YoudinGoodman2005}. Schematically, in the linear phase, dust is dragged toward pressure perturbation maxima, and gas is dragged along, with the help of the Coriolis force, strengthening the pressure maxima, hence a positive feedback \citep{Jacquetetal2011SI} (see also \cite{SquireHopkins2020}). In the nonlinear phase, an overdensity of pebbles drags the surrounding gas along, collectively moving like a much larger body and collecting isolated pebbles which move with different radial and azimuthal velocities. This way, the initial overdensity grows. Numerical models show that, under some conditions, the dust structures produced by the instability fragment into very dense clumps which reach the Roche density and gravitationally collapse \citep[][and many others]{Johansenetal2007,Simonetal2016SI, Yangetal2017}, as presented in Fig.~\ref{fig:SI}. 

\begin{figure}[t]
  \centerline{\includegraphics[width=0.8\textwidth]{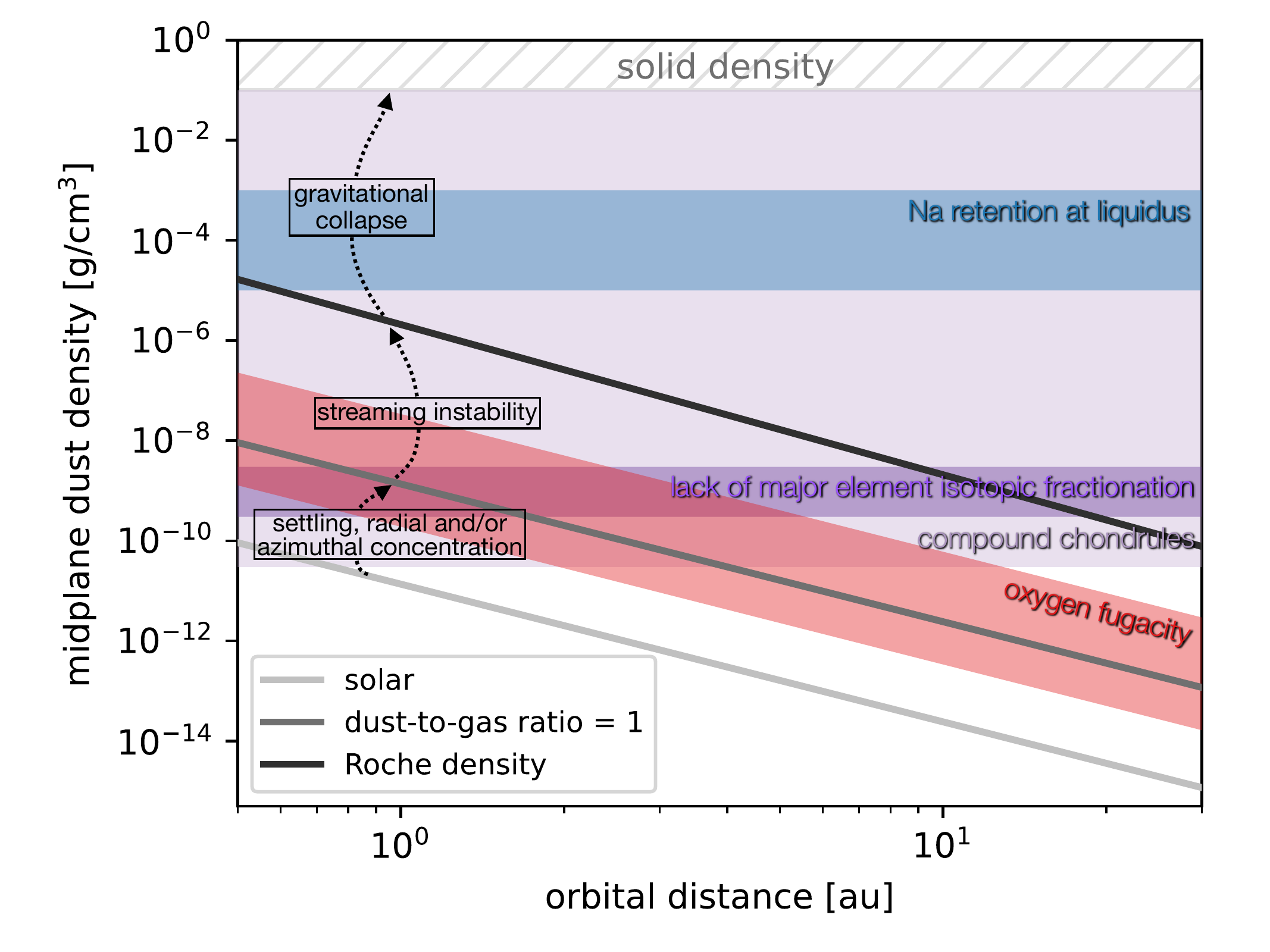}}
  \caption{\label{fig:denschondr} Comparison of typical midplane dust densities in the Minimum Mass Solar Nebula model reached during various stages of planetesimal formation (see the text boxes) and the constraints on chondrule formation derived from: the oxygen fugacity estimated from olivine Fa contents (assuming a CI anhydrous dust$\pm$ice composition for the solids; \cite{Tenneretal2015}), lack of major element isotopic fractionation \citep{CuzziAlexander2006}, compound chondrules (assuming a 1 h collision time; \cite{Jacquet2021}), Na retention at liquidus in type II chondrules \citep{Alexanderetal2008}. We note that the oxygen fugacity constraint is a constraint on the solid/gas ratio so is proportional to the assumed density profile; in showing the total range from type I to type II chondrules irrespective of the heliocentric distances, we have also not tried to refine the constraints as a function of their presumed provenance, e.g. with respect to the snow line (both chondrule types occur in the two meteorite superclans anyway). We also note that the compound chondrule constraint is only a lower limit, because we only bound the chondrule-chondrule relative velocity by a \textit{maximum} of 1 km/s \citep{ArakawaNakamoto2019}.}
\end{figure}

  Since the streaming instability would operate rapidly (within a few hundred or thousand orbits, depending on St; \citet{Yangetal2017}), any excess solid abundance above the threshold for its operation would rapidly be locked up in new planetesimals and the background disk dust density would return to marginal stability. Hence, one might expect a streaming instability-regulated solid/gas ratio around unity. This compares favorably with many chondrule constraints plotted in Figure~\ref{fig:denschondr}, which are consistent for the inner part of the solar nebula, up to a few au. Still, the incompatibly much higher densities (above the Roche density) estimated by \citet{Alexanderetal2008} from Na retention at liquidus temperatures for type II chondrules, would need to be explained away somehow (e.g. if the Na-bearing olivine cores, which are normally associated with the onset of olivine crystallization, are in fact relict grains). So chondrules may have formed in the settled dust disk, but not during the individual collapse events therein, to which we now turn.

\subsection{Collapse of the pebble cloud}
%How rapid is the accretion of a single planetesimal? [single chondrite groups appear homogeneous in composition across the various putative depths of provenance]
Once the streaming instability leads to very dense, gravitationally-bound pebble clumps, the following collapse to planetesimals is thought to be very quick. In this sense, the planetesimals produced this way may be seen as a snapshot of the local disk composition. This seems to be consistent with the fact that single chondrite groups appear homogeneous in bulk isotopic composition even though their members may come from different depths within their parent body \citep{Pedersenetal2019, Przylibskietal2023}. 

\citet{WahlbergJanssonJohansen2017} and \citet{Visseretal2021} studied the collapse of pebble clouds with one-dimensional models, assuming they are spherical and not rotating but including collisions between the constituent pebbles. They found that in the presence of protoplanetary gas, the collapse only takes about one-third of the local orbital time. Both models agree that there should be some degree of dynamical sorting of the infalling pebbles, such that the largest pebbles end up closest to the final planetesimal center and the smallest pebbles at its surface. This is qualitatively consistent with the sizes of CO chondrite chondrules increasing with petrological subtype and thus depth inside of the parent body \citep{Pintoetal2021}. However, in the models cited above, only one type of material was included and the pebbles might be dusty agglomerates rather than individual chondrules or CAIs, whose admixtures are yet to be studied in this context. 

\citet{Nesvornyetal2021} and \citet{PolakKlahr2023} followed the pebble cloud collapse in three-dimensional models, including possible initial rotation of the pebble cloud. In these models, the collapse takes slightly longer than in the one-dimensional models but the timescales are still comparable with the local orbital time. Remarkably, a significant amount of binary planetesimals are formed with predominantly prograde rotation, which is consistent with the observations of Trans-Neptunian Objects \citep{Fraseretal2017,Nesvornyetal2019}. However, these three-dimensional models assumed monodisperse pebbles, thus they could not study the possible size sorting during the collapse.

\subsection{Planetesimal formation sites}

  Although streaming instability could, indeed, take over for solid/gas ratios of order unity at the midplane, this is not an altogether easy threshold to reach. The hydrodynamic models, such as presented in Fig.~\ref{fig:SI}, are done in very local domains due to the very high resolution needed to resolve the streaming instability, and thus assuming arbitrary initial conditions. Global models are needed to study the interplay of dust coagulation and radial drift potentially leading to streaming instability conditions at some locations in the disk. However, often such global models of dust evolution do not lead to any planetesimal formation at all \citep{EstradaUmurhan2023}. The very pressure bumps mentioned above, which concentrate solids radially, may be the necessary havens for accretion \citep{Stammleretal2019,JiangOrmel2021,Lauetal2022}. Several models have shown that planetesimal formation is favored in the vicinity of the water snowline \citep{SchoonenbergOrmel2017,DrazkowskaAlibert2017,Charnozetal2019,Hyodoetal2019,Hyodoetal2021} or the silicate sublimation line \citep{Morbidellietal2022,Izidoroetal2022}. Independently of pressure bumps, their solid abundances could be indeed increased there because evaporation would slow down the inward drift of condensable matter inside them, by zeroing the drift relative to gas (which may have originally flowed outward; \cite{Morbidellietal2022}) for the evaporated species, and also reducing the size of the remaining, more refractory dust grains \citep[e.g.][]{CieslaCuzzi2006}. So mass would accumulate more rapidly from the outer disk than it could be removed.

%  The pressure bump could cut off the supply of volatile ices so, if it remains fixed (e.g. if maintained by a giant planet), nothing would condense past the receding ice line (as viscous heating diminishes), and thus ices would be confined outside the original (hereby "fossilized") position of the ice line \citep{Morbidellietal2016}, hence perhaps the paucity of ice-rich bodies inside 2.5 au although the water snow line of a passive disk around a proto-Sun should lie near 1 au \citep[e.g.][]{ChiangGoldreich1997}.

  From the meteorite record, we also know that planetesimal accretion started early after CAI formation, which possibly corresponds to the protoplanetary disk buildup stage, and proceeded for several Ma. This can be generally reproduced in models following disk formation and modeling the planetesimal formation at the snow line \citep{DrazkowskaDullemond2018,Lichtenbergetal2021}. However, fulfilling all the requirements at once: that is starting early, forming planetesimals at more than one location, and keeping pebble supply for continuing planetesimal formation over several Ma, is very challenging \citep{MarschallMorbidelli2023}. One possibility is photoevaporation increasing the solid/gas ratio at the midplane \citep{Carreraetal2017,Ercolanoetal2017}, which may explain a new surge in planetesimal accretion after a relative dearth between 1 and 2 Ma after CAIs \citep{SugiuraFujiya2014}, as proposed by A. Morbidelli in the workshop. Another would be to include the full planet formation model in the planetesimal formation models. To date, these models typically only consider dust growth to pebbles and planetesimal formation but without their further evolution. Nevertheless, it is known that a giant planet, if formed early, may trigger another burst of planetesimal formation outside of its orbit \citep{Erikssonetal2020,ShibaikeAlibert2020,Milleretal2021}, consistent with the concept of Jupiter splitting the NC and C reservoirs. 
%When/where can SI (or other possible accretion mechanisms) operate? [from Hf-W dating et al., planetesimal accretion started early, in both NC and C superclans, went on for a few Ma]

\subsection{Rapid formation of giant planets?}

Certainly, as we saw in the previous section, the rapid formation of a giant planet may prevent pebbles from drifting inward too quickly and to keep planetesimal formation going for several Ma, consistent with the Solar System record.  However, such a fast core growth is not typically seen in planet formation models. The timescale of Jupiter formation is a long-standing problem in planet formation theory, first recognized by \citet{Pollacketal1996}. In the classical models, where planet growth is driven by the accretion of planetesimals, already forming Jupiter rocky core before the gas disk dissipates is a challenge. 

  An alternative could be top-down formation of the giant planet by fragmentation of gravitationally unstable disk \citep{Boss1997,Mayeretal2002}. However, many problems have been pointed out regarding this mode of planet formation. The protoplanetary disk would need to be atypically cold to allow for its fragmentation \citep{Kratteretal2010}. Planets forming this way would be much more massive than Jupiter and on wider orbits but migrate quickly in those massive disks and fall onto the central star \citep{KratterLodato2016}. What is most important for us, the early planetesimal formation is likely hampered in massive gravitationally unstable, turbulent disks, as streaming instability requires a calm midplane pebble layer \citep{DrazkowskaDullemond2018} and the spiral waves present in gravitationally unstable disks are not helping to enhance the pebble-to-gas ratio \citep{Vorobyovetal2018}. Further studies are needed on the perspective of planetesimal formation in gravitationally unstable disks and its interplay with planet formation in this scenario. 

In recent years, the \textit{pebble accretion} paradigm was introduced, where planetary core growth is aided by the accretion of the pebbles rather than planetesimals. Pebble accretion may be faster as pebbles are not scattered by the gravity of massive core as easily as planetesimals \citep{OrmelKlahr2010,LambrechtsJohansen2012}. However, even with pebble accretion, the formation of Jupiter's core in a smooth disk with the initial dust-to-gas ratio of 0.01 takes over 1 Ma \citep{LambrechtsJohansen2014}. This result is very sensitive to the exact model parameters such as the turbulence level and disk mass \citep{Drazkowskaetal2021}. It is worth noting that in order to accrete pebbles efficiently, the planetary core must already be quite massive with the minimum mass sensitively dependent on the disk pressure gradient and pebble Stokes number \citep[see, e.g.,][]{LiuJi2020}. At 1~au in a typical disk model, this mass would correspond to 100~km-sized body and it increases with the orbital distance. 

 It seems that, within the core accretion paradigm, the only way to produce massive planetary cores quickly is to start with an initial pressure bump helping to enhance the solid abundance up to the streaming instability threshold and at the same time providing an abundance of pebbles for fast pebble accretion \citep{Morbidelli2020,JiangOrmel2023}. \citet{Lauetal2022} presented a model in which a core of several Earth masses forms from a pebble ring at 10~au on the timescale of just 0.1 Ma. This is possible thanks to considering pebble growth and drift, planetesimal formation, and planetesimal growth by both planetesimal and pebble accretion simultaneously. This way, a few massive bodies emerging out of the very first planetesimals grow quickly and stir the planetesimals formed later preventing the emergence of further planetary cores, which otherwise would compete for accreting pebbles present in the ring. So the pressure bump invoked for the NC/C dichotomy \citep{BrasserMojzsis2020} would be a cause rather than an effect of the growth of Jupiter. Now, while the observation of the NC/C dichotomy for early accreted iron meteorites is generally attributed to an early growth of Jupiter \citep{Kruijeretal2017}, the forming Jupiter may have efficiently accreted bodies with intermediate isotopic compositions \citep{%Stammleretal2019,
Jacquetetal2019}, or that intermediate region of the disk may have been unfavorable to planetesimal formation, if devoid of pressure bumps \citep{Morbidellietal2022}.

\section{Conclusion}\label{sec13}
As resolution of astronomical observations improves, our picture of the solar nebula may become less nebulous. Rings such as those observed in neighbouring protoplanetary disks may have been the safe havens for preventing drift loss of solids and producing age-diverse collections of them as observed in chondrites. These would have been concentrated to the level recorded by chondrules, conducive to the streaming instability and thence planetesimal accretion. The rings could interrupt, or limit, communications between different disk regions (e.g. inner and outer disk) accounting for the NC/C isotopic dichotomy among meteorites. Presumably, these rings trace pressure maxima, however the origin of such pressure maxima remains contentious. Giant planets such as Jupiter might be responsible, but such planets may also be consequences of such pressure maxima. In early times, more or less contemporaneously with infall, large-scale transport may have been unimpeded, as suggested by refractory inclusions, now chiefly found far from the Sun. None of this, of course, makes a standard model of meteorite formation "geography" yet, but this at least offers tantalizing hints to explore further, whether in the microscope or the telescope, or the minds above them.

\backmatter

\bmhead{Acknowledgments}
The authors are grateful to the organizers of the International Space Science Institute workshop "Evolution of the Solar System: Constraints from Meteorites" (June 5-9, 2023, in Bern) and the two anonymous reviewers of this chapter. JD was funded by the European Union under the European Union’s Horizon Europe Research and Innovation Programme 101040037 (PLANETOIDS). This chapter is dedicated to the memory of the lead author's grandmother Marguerite Jacquet (1923-2023). \textit{Nous aurons bien ri de l'\'{e}ternit\'{e}}.

\section*{Declarations}

\paragraph{Competing interests}
The authors have no competing interests to declare that are relevant to the content of this article.

\paragraph{Data availability} 
Not applicable.


%\bibliographystyle{natbib}%{apj}%{aa}

\bibliography{bibliography.bib}% common bib file


\end{document}